\documentclass[conference]{IEEEtran}

\usepackage{cite}

\ifCLASSINFOpdf
 \usepackage[pdftex]{graphicx}

\else

\fi

\begin{document}

\title{A fourth explanation to Brooks' Law\\ --- The aspect of group developmental psychology}

\author{\IEEEauthorblockN{Lucas Gren}
\IEEEauthorblockA{Chalmers University of Technology and \\The University of Gothenburg\\
Gothenburg, Sweden 412--92\\
Email: lucas.gren@cse.gu.se}

}

\maketitle

\begin{abstract}
Brooks' Law is often referred to in practice and states that adding manpower to a late software project makes it even later. Brooks' himself gave three explanation only related to concrete task-related issues, like introducing new members to the work being done, communication overheads, or difficulty dividing some programming tasks. Through a description of group developmental psychology we argue for a fourth explanation to the law by suggesting that the group will fall back in its group development when new members are added, resulting in rework setting group norms, group goals, defining roles etc.\ that will also change over time. We show that this fourth explanation is important when trying to understanding Brooks' Law, and that adding the group developmental perspective might help software development organizations in managing projects.
\end{abstract}

\IEEEpeerreviewmaketitle

\section{Introduction}
In 1975 Frederick Brooks came out with his famous book ``The Mythical Man-Month: Essays on Software Engineering''~\cite{brooks1975tmm}. The idea that adding manpower to a late software project makes it even later, became well-known in the software engineering field, and Brooks himslef called his idea Brooks' Law. Brooks gave the following three explanations to his law: (1) Ramp-up time to get productive (getting introduced to the work already conducted). (2) Communication overheads (more people means more people to communicate with). (3) Limited divisibility of tasks (some tasks cannot be divided).

All these explanations focus on the actual task to be solved, however, there are also social-psychological factors that change and develop over time in work-groups that provide an additional explanations to the law, namely group developmental aspects from a psychological perspective. 

We will now first present what we mean by ``groups'' and ``teams,'' present an integrated model of group development followed by the three explanations given by Brooks in more detail, discuss group developmental connections to Brooks' Law, and finally conclude and suggest future work.

\subsection{Groups and Teams}
According to \cite{grupp}, a group can be defined as ``three or more members that interact with each other to perform a number of tasks and achieve a set of common goals.'' A group that is larger might not have a common goal for all group members and then might, in fact, instead consist of subgroups. Some studies have shown that group around eight individuals are the most effective~\cite{wheelan2009}. Sometime a work-group and a team are separated by the fact that a team has found effective means to achieve its goals, unlike the work-group. However, the terms are used somewhat interchangeably in this paper, since software engineering work-groups are most often called ``teams'' no matter their actual effectiveness.

Due to many years of research on groups in social and organizational psychology, there are a diversity of group development models \cite{wheelan1993}. Even though the models have their differences, there seems to be a reoccurring pattern of what happens to all types of groups when humans get together in order to solve a task together. The first researcher to integrate group models into a general group development theory over time was Tuckman~\cite{tuckman} in 1965. In the nineties Susan Wheelan conducted a similar aggregation that resulted in the Integrated Model of Group Development. However, Tuckman's~\cite{tuckman} model with the phases; Forming, Storming, Norming, and Performing can, for the most part, be translated into the stages as suggested by Wheelan \cite{wheelan}. These four phases will be presented next.

\subsection{Wheelan's Integrated Model of Group Development}\label{sub:integratedgroup}
\begin{figure*}
\centerline{\includegraphics[width=130mm]{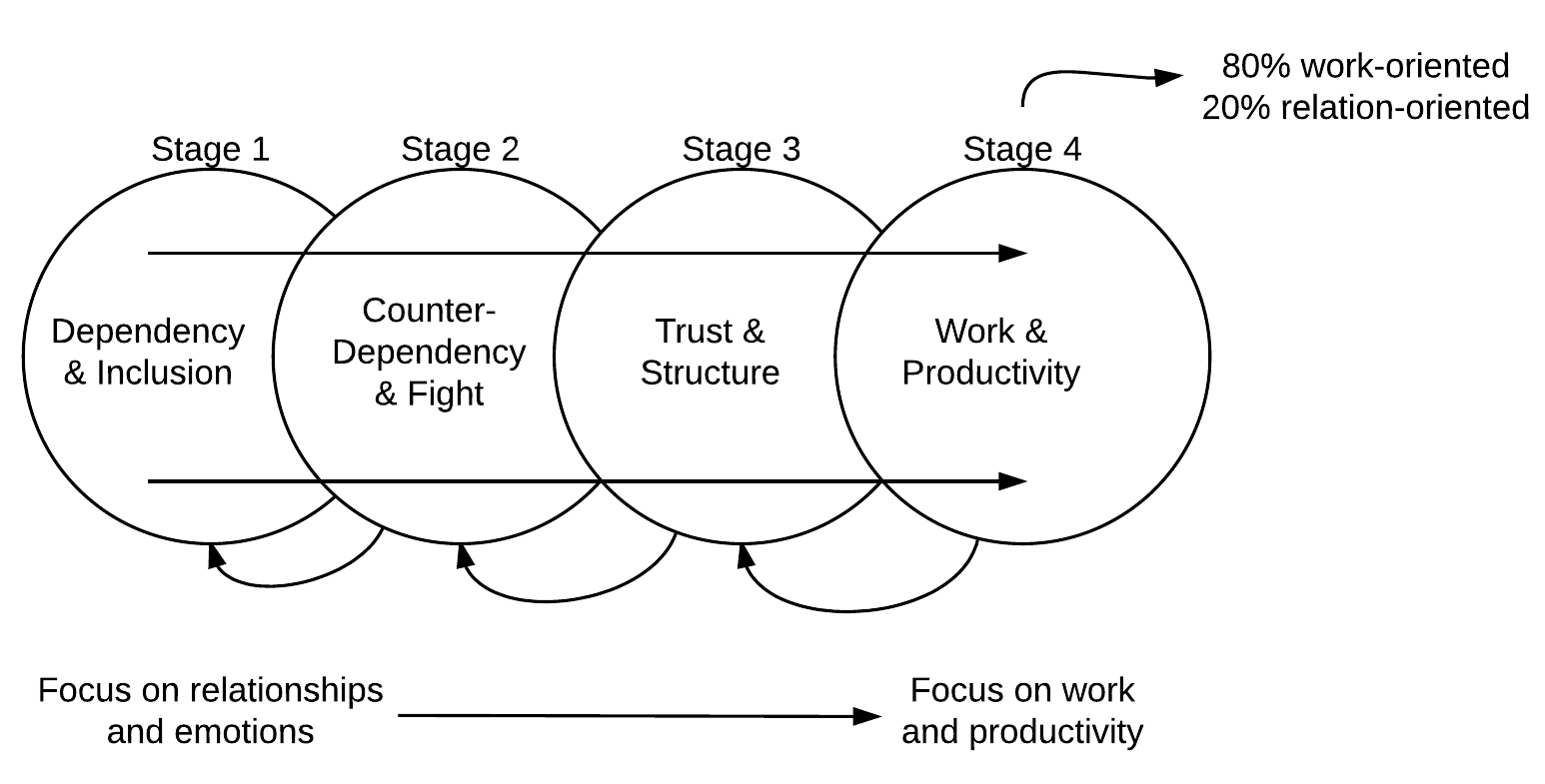}}
\caption{The Group Development Stages. Adopted from~\cite{wheelan2012}.}
\label{fig:groupstages}
\end{figure*}

The Integrated Model of Group Development (IMGD for short) presents four different temporal stages that all groups go through on their journey to becoming a mature high performing team. These stages are shown in Figure~\ref{fig:groupstages} and are described next.

\paragraph{Stage 1 --- Dependency and Inclusion}
In the first stage, the group members have focus on safety and inclusion, a dependency on the designated leader, and more of a wish for order and structure, than in later stages. A work-group in the earlier stages can still get work done, but the members will focus more on figuring out who the other members are and need time to feel safe in the group and get to focus more on work. In a stage one group, there is an evident lack of structure and the group needs to figure out how to get organized in order to eventually, or hopefully, conduct efficient work and reach the group's goals. The newly formed groups will lack a sense of belonging and team spirit, which the groups needs to work on achieving (i.e.\ they are simply not a team yet). Directive leadership is expected and appreciated by the group members since it contributes to the feeling of safety. In stage one, the group tends to have had role division based on superficial status and not real competence, since the group members simply did not know each others' skills yet. When the group feels comfortable enough the member will start questioning goals, roles, and the organization of work, which is the next stage of group development~\cite{wheelan}.

\paragraph{Stage 2 --- Counter-Dependency and Fight}
When the group members have been given some time to figure each other out, they will experience an increased feeling of safety. With that safety comes the will of getting the work done more efficiently, which means that the group members start questioning their own and others' roles as well as the group goals and the organization of work. In order to find more efficient ways of working, the group starts having conflict. However, such conflicts are constructive when work-related and is a must in order to organize and make use of the real competences of the group members in a better way. These more turbulent times build trust and cohesion in the group and is unavoidable~\cite{wheelan}.

\paragraph{Stage 3 --- Trust and Structure}
After a period of questioning each other and the leader a better work structure is starting to form in the group. The roles are based on actual competence, the leader needs to be less directive, and as the group matures it is also ready to self-organize more and more. The communication patterns will be more diverse and more task-orientated, meaning that the group members talk to whoever they need from a work-related perspective. There is a more evident consensus on work goals, simply because the group members discussed them thoroughly in the second stage. Conflict will still occur, but the difference is that the group has experience of solving conflict if reaching this stage, which means that the conflicts tend to be managed more effectively and be solved much faster~\cite{wheelan2012}.

\paragraph{Stage 4 --- Work and Productivity}
The fourth stage is more and better division of work, role division, etc. At this stage the productivity is much higher and the work-group turns into what is often referred to as a high performing team. The group cohesion is high and so is also the inter-personal attraction between team members. Stage four groups are highly effective and also evaluate and question their work methods and make sure to assess the quality of their output. Both self-organization and continuous improvement are characteristics of mature groups from a group dynamics perspective. It is also a fact that many factors within the organization or group can push the group back to earlier developmental stages. All external and internal changes will result in the work group having to redo setting norms and negotiating roles, goals, etc. Basically, all changes will have such an effect, e.g.\ change of demands from the organization, losing staff, getting new staff, and so on and so forth. Another characteristic of a high performing team is that decision-making is participatory because all members are needed in order to achieve the team goals, the team must not have too few or to many members. Getting to stage four takes a lot of work both from the group members but the group also needs to be given the right possibilities from their surrounding ecosystem~\cite{wheelan2012}.

\paragraph{Measuring Group Development}\label{group}
Wheelan~\cite{wheelan2012} was not the first researcher who suggested these characteristic stages of group development, but she developed a questionnaire that can be used to measure these different stages with four scales. Her tool has made it possible to measure and diagnose where a specific group is focusing its energy from a group developmental perspective. The scales have been shown to correlate with a diversity of effectiveness and productivity measurements in different fields. The scale that measures stage four (``work and productivity'') and has been shown to correlate with an ability to finish projects faster~\cite{wheelan1998}, better student performance on standardized test (SAT scores), if the faculty team scores high on GDQ4~\cite{wheelan1999,wheelan2005}, and the fact that intensive care staff save more lives in surgery~\cite{wheelan20032}.


\subsection{Brooks' Law and the Three Given Explanations}
According to Brooks himself, the law can in general be explained by three factors~\cite{brooks1975tmm}. The first factor is that new team members need time to learn what the team is doing. Brooks writes: ``Each worker must be trained in the technology, the goals of the effort, the overall strategy, and the plan of work.'' p.\ 18~\cite{brooks1975tmm}. He also mentions team organization, task division and that software engineering often includes complex endeavors and the new member needs to essentially get educated in the work being done. This means that the other team members must be the educators and the team will then lose productivity overall. Also, while the new member is learning they might as well introduce bugs etc.\ that also means more work for the team.  As mentioned in the introduction, Brooks calls this ``ramp-up'' time. The second aspect is what Brooks calls additional communication overhead. Having more people work on the same task, they must all communicate and synchronize their work continuously, which also decreases the overall productivity. The third and last explanation given is the fact that some tasks can not be divided easily. Then, adding more team members will have no effect in task completion time.

\section{Discussion}\label{sec:disc}
If groups have been shown to fall back in their development when new members are added to work groups \cite{wheelan}, this indicates that a fourth explanation to Brooks' Law would be the fact that the group as a whole does, not only need to introduce the new member to the actual work being done from a work content perspective, but also needs to more or less implicitly, introduce the new member(s) into the group norms, roles, goals, and how work is organized from a psychological perspective. We know from group developmental psychology that such a process does not happen instantaneously, and that a new group member needs to iterate through the stages of Dependency \& Inclusion, Counter-Dependency \& Fight, Trust \& Structure, and Work \& Productivity. This will happen faster or slower depending on how well the new team member knows the rest of the team, and how well the other team members include the new member. However, if the new developer has never met the rest of the team, this person must first figure out who the other group members are, then dare to question their and the person's own roles, question the goals and the organization of work, learn how conflicts are manged, and so on and so forth, and finally being fully integrated as a member of a potentially high performing team (if the team was at that stage when the new resource was added). The team members must spend time including the new member from a psychological perceptive if they want a new resource that can actually contribute to the work being done. However, if the team is under time pressure such ``inclusion work'' might just as well be futile since there is not enough time to introduce group norms to a new resource since they are implicit and have to be experienced. 


All-in-all, we believe Brooks' Law can then be explained by the follow four points:

\begin{enumerate}
\item Ramp-up time to get productive (getting introduced to the work already conducted).
\item Communication overheads (more people means more people to communicate with).
\item Limited divisibility of tasks (some tasks cannot be divided).
\item The group will fall back in its group development (the new resource will need time, and take time, from the other developers when learning the rules of the game (i.e.\ the group norms etc.) from a psychological perspective).  
\end{enumerate}

Brooks probably observed group developmental issues, but did not mention them explicitly. In his explanation of ramp-up time he explicitly mentions ``team organization'' which, of course, includes aspects of how that specific team works together. However, the aspect of that teams will fall back in their group development and therefore be less productive from a group developmental perspective is not mentioned.

We found very few researchers that have investigated Brooks' Law in more detail. The only scientific article found was \cite{hsia1999brooks} and showed that if resources are added enough in advance, the software projects will not be late, but instead saved in the intended way. According to \cite{mccain2006influential}, Brooks' Law was continuously cited up until year 2000 and we have not been able to find more recently published citation statistics. Also, in \cite{verner199925years} it was concluded that the aspects pointed out by Brooks were still the main causes of project failure. 

In a well-referenced non-scientific source, Berkun \cite{berkun} provides a list of exceptions to Brook's Law: 

\begin{enumerate}
\item It depends who the manpower is (Berkun \cite{berkun} stated that he would consider adding a programmer late if the person knows the code base and is friends with half of the team).
\item Some teams can absorb more change than others (meaning some teams can include and teach new members more efficiently).
\item There are worse things than being later (which means that being late might be acceptable if high quality is reached for example). 
\item There are different ways to add manpower (meaning that management can introduce new resources in a way that their roles are explained and clarified to the team, which would help the integration of new members).
\item It depends on why the project was late to begin with (if the project was late because of poor team practices new expertise might help).
\item Adding people can be combined with other management action (removing a poor programmer and team player and add an excellent one might be a very good idea for the project, according to \cite{berkun}). 
\end{enumerate}
We believe many of these points can be explained from a group developmental perspective, and then, do not have to be seen as exceptions to the law, but instead be explained by our fourth explanation. We still believe, however, that the law holds, but the impact of adding resources late can be of different magnitude. A reference to group development psycholgogy for each given exception is now presented. (1) Adding a resource that is friends with half of the team would make the team integration of that resource much faster, and then might be worth it if there is enough time left until the deadline, while an unknown resource would not be worth adding. (2) It is well-known, even in software engineering \cite{grenjss2}, that both technical and group-working skills are needed for a team to function well. A team that is defined as including members with excellent social skills, it might be worth it adding a more unknown resource, even at a later stage. (3) The third point defined by Berkun \cite{berkun} is simply the idea of letting a project be late. (4) The fourth point is also about doing a better job as managers when adding resources then simply letting the team figuring it out themselves. Such strategies are connected to group development since they aim at helping the team integrate the new resource faster from a psychological perspective (since management often do not have the knowledge to introduce the code base to the new resource for example). (5) If a team is at the more immature stages of group development, adding a resource that can provide order and structure to the work being done, might increase the productivity if time allows for such a development. (6) We know from social psychology that one person's negative attitude is infectious and can more or less ruin the teamwork (see e.g.\ \cite{barsade2002ripple}), the team cohesion, and therefore hinder the team from performing, which means getting stuck in the more immature stages of group development. Removing such an individual and replacing him or her would then be a good idea. However, such a decision is hard for a manger to make since one needs to be certain of that the groups' problems are due to only one individual, which is seldom the case. Even if it appears to be so, scapegoating, and other social psychological behavior of groups much first be disproved. A wrong decision might have devastating consequences to the individual and the team. If the team has gotten stuck in the earlier stages of group development an intervention to clarify goals, roles, and solve conflict might be more appropriate, if time allows.

Opelt \cite{opelt} claims that because of being aware of Brooks' Law, software development somewhat changed, and that teams that were using XP practices (mainly collocation and pair programming) in the first decade of the third millennium, fixed many of the issues that lead to Brooks' Law. That study is a short research paper based on experiences from one team and is therefore very difficult to generalize from. However, what is clear in recent research is that the concept of an ``agile team'' is connected to what we mean by a high performing team in group psychology \cite{grenjss2}. Even in the most modern agile teams especially our added fourth explanation to the law will still be valid, since even more team-based worked is imposed. No matter if we have a perfect agile team where everybody knows everything about the project and can switch roles at no overhead costs, Brooks' Law would still hold. Maybe the reader notices the irony, and yes, we believe having different roles and difference expertise in teams are \emph{de facto} always the reality, and since we believe aspects of Brooks' Law is due to group developmental issues, we believe we can still improve our management of software development teams, and make better evidence-based decisions if the psychological aspects of teams are also well understood.

\section{Conclusion and Future Work}
This paper set out to provide a fourth explanation to Brooks' Law by adding the aspect of group developmental psychology. Through describing an integrated model of group development and connecting it to Brooks' given explanations, we have shown that our fourth explanation is relevant to understanding Brooks' Law. These findings are important contributions to software development organizations and managers dealing with delayed software projects. In terms of future research, we particularly suggest empirically investigating measurements of group development and their relation to adding resources to projects late in the project life-cycle.

\bibliographystyle{IEEEtran}

\bibliography{references}

\end{document}